\documentclass[longauth]{aa}
\usepackage{graphicx}
\usepackage{txfonts}
\usepackage{epsfig}
\begin{document}
   \title{AGILE detection of a rapid $\gamma$-ray flare from the
  blazar PKS 1510-089 during the GASP-WEBT monitoring\thanks{The radio-to-optical data presented in
  this paper are stored in the GASP-WEBT archive; for questions regarding
  their availability, please contact the WEBT President Massimo Villata.}}

\author{F.~D'Ammando\inst{1,2}, G.~Pucella\inst{1}, 
  C.~M.~Raiteri\inst{3}, M.~Villata\inst{3}, V.~Vittorini\inst{1,4}, S.~Vercellone\inst{5},
  I.~Donnarumma\inst{1}, F.~Longo\inst{6}, M.~Tavani\inst{1,2}, A.~Argan\inst{1}
% ...
} 

\institute{$^{ 1}$ INAF/IASF--Roma, Via Fosso del Cavaliere 100, I-00133 Roma, Italy \\  
           $^{ 2}$ Dip. di Fisica, Univ. di Roma ``Tor Vergata'', Via della
           Ricerca Scientifica 1, I-00133 Roma, Italy \\   
           $^{ 3}$ INAF, Oss. Astronomico di Torino, Via Osservatorio 20, I-10025 Pino Torinese
           (Torino), Italy \\
           $^{ 4}$ CIFS--Torino, Viale Settimio Severo 3, I-10133 Torino, Italy \\ 
           $^{ 5}$ INAF, Istituto di Astrofisica Spaziale e Fisica Cosmica,
           Via U.~La Malfa 153, I-90146 Palermo, Italy
           \\
  $^{ 6}$ Dip. di Fisica and INFN Trieste, Via Valerio 2, I-34127 Trieste, Italy \\ 
           $^{ 7}$ INFN--Pavia, Via Bassi 6, I-27100 Pavia, Italy \\ 
           $^{ 8}$ Dip. di Fisica Nucleare e Teorica, Univ. di Pavia, Via
           Bassi 6, I-27100 Pavia, Italy \\
           $^{ 9}$ INAF/IASF--Bologna, Via Gobetti 101, I-40129 Bologna, Italy \\ 
           $^{10}$ INAF/IASF--Milano, Via E.~Bassini 15, I-20133 Milano, Italy
           \\
           $^{11}$ Dip. di Fisica Generale dell'Universit\'a, Via P.~Giuria 1, I-10125 Torino, Italy
           \\
           $^{12}$ ENEA--Bologna, Via dei Martiri di Monte Sole 4, I-40129 Bologna, Italy \\ 
           $^{13}$ INFN--Roma ``La Sapienza'', Piazzale A. Moro 2, I-00185
           Roma, Italy \\
           $^{14}$ INFN--Roma ``Tor Vergata'', Via della Ricerca Scientifica 1, I-00133 Roma, Italy \\
           $^{15}$ INAF--Oss. Astronomico di Cagliari, loc. Poggio dei Pini,
           strada 54, I-09012 Capoterra (CA), Italy \\ 
           $^{16}$ Dip. di Fisica, Univ. dell'Insubria, Via Valleggio 11, I-22100 Como, Italy \\ 
           $^{17}$ ENEA--Roma, Via E. Fermi 45, I-00044 Frascati (Roma), Italy
           \\ 
           $^{18}$ Instituto de Astrof\'{\i}sica de Andaluc\'{\i}a, CSIC, Apartado 3004,
18080 Granada, Spain \\
           $^{19}$ Department of Astronomy, University of Michigan, Ann Arbor,
           MI 48109, USA\\
           $^{20}$ Pulkovo Observatory, Russian Academy of Sciences, 196140, St.-Petersburg, Russia \\         
           $^{21}$ MPIfR, D-53121 Bonn, Germany \\
           $^{22}$ Instituto de Astronomia, Universidad Nacional Autonoma de
           Mexico, Mexico, D. F. Mexico \\
           $^{23}$ Tuorla Observatory, Department of Physics and Astronomy, University of Turku, V\"ais\"al\"antie 20, 21500 Piikki\"o, Finland \\
           $^{24}$ Astron.\ Inst., St.-Petersburg State Univ., 198504
           St.-Petersburg, Russia \\
           $^{25}$ INAF--Osservatorio Astrofisico di Catania, Via S. Sofia
           78, I-95123 Catania, Italy\\ 
           $^{26}$ Institute of Astronomy, National Central University,
           Taiwan\\
           $^{27}$ Lulin Observatory, Institute of Astronomy, National Central
           University, Taiwan \\ 
           $^{28}$ INAF, Osservatorio Astronomico di Roma, Via di Frascati 33,
           I-00040, Monte Porzio Catone, Italy \\ 
           $^{29}$ INAF, Osservatorio Astronomico di Collurania, Via Mentore
           Maggini, I-64100 Teramo, Italy\\
           $^{30}$ Harvard-Smithsonian Center for Astrophysics, Cambridge,
           Garden st. 60, MA 02138, USA \\
           $^{31}$ Institute for Astrophysical Research, Boston University, 725
Commonwealth Avenue, Boston, MA 02215, USA \\
           $^{32}$ ZAH, Landessternwarte Heidelberg, K\"onigstuhl, D-69117
           Heidelberg, Germany \\
           $^{33}$ Instituto de Astronomia, Universidad Nacional Autonoma de Mexico,
           2280 Ensenada, B.C. Mexico \\
           $^{34}$ Abastumani Observatory, 383762 Abastumani, Georgia \\
 $^{35}$ INAF--IRA, contrada Renna Bassa, I-96017 Noto (SR), Italy \\
 $^{36}$ Department of Physics and Astronomy, University of Southampton S017 1BJ, UK \\
          $^{37}$ INAOE, Apdo. Postal 51 y 216, 72000 Tonantzintla,
Puebla, Mexico \\
 $^{38}$ Lowell Observatory, Flagstaff, AZ 86001, USA \\
           $^{39}$ ASI--ASDC, Via G. Galilei, I-00044 Frascati (Roma), Italy \\
           $^{40}$ ASI, Viale Liegi 26, I-00198 Roma, Italy \\ 
   } 
           
\offprints{F. D'Ammando, \email{filippo.dammando@iasf-roma.inaf.it} }

  \date{received; accepted}

 
  \abstract
%
   {We report the detection by the AGILE satellite of a rapid $\gamma$-ray
  flare from the source 1AGL J1511$-$0908, associated with the powerful
  $\gamma$-ray quasar PKS 1510$-$089, during a pointing centered on the Galactic
  Center region from 1 March to 30 March 2008. This source has been continuosly
  monitored in the radio-to-optical bands by the
   GLAST-AGILE Support Program (GASP) of the Whole Earth Blazar Telescope
   (WEBT). Moreover, the $\gamma$-ray flaring episode triggered three ToO observations by the $Swift$
   satellite in three consecutive days, starting from 20 March 2008.}
%
   { The quasi-simultaneous
   radio-to-optical, UV, X-ray and $\gamma$-ray coverage allows us to study in
   detail the multifrequency time evolution, the spectral energy distribution of this source and its theoretical
   interpretation based on the
   synchrotron and inverse Compton (IC) emission mechanisms.}
%
   {During the radio-to-optical monitoring provided by the GASP--WEBT, AGILE observed the source with its two co-aligned imagers, the Gamma-Ray
   Imaging Detector (GRID) and the hard X-ray imager (SuperAGILE),
   sensitive in the 30~MeV--30~GeV and 18--60 keV energy bands,
   respectively.}
%
   {In the period 1--16 March 2008, AGILE detected $\gamma$-ray emission from PKS 1510$-$089 at a significance level of
   6.2-$\sigma$ with an average flux over the entire period of $(84 \pm 17) \times 10^{-8}$ photons
   cm$^{-2}$ s$^{-1}$ for photon energies above 100 MeV. After a predefined satellite
   re-pointing, between 17 and 21 March 2008, AGILE detected the source at a
   significance level of 7.3-$\sigma$, with an average flux (E $>$ 100 MeV) of
   $(134 \pm 29) \times 10^{-8}$ photons cm$^{-2}$ s$^{-1}$ and a peak level
   of $(281 \pm 68) \times 10^{-8}$ photons cm$^{-2}$ s$^{-1}$ with daily integration. During the
   observing period January--April 2008, the source also showed an intense and
   variable optical activity, with several flaring episodes and a significant
   increase of the flux was observed at millimetric frequencies. Moreover, in the X-ray band the $Swift$/XRT
   observations seem to show an harder-when-brighter behaviour of the source spectrum.} 
%
   {The flat spectrum radio quasar PKS 1510$-$089 showed strong activity between
   January and April 2008, with episodes of rapid variability
   from radio to $\gamma$-ray energy bands, in particular with a rapid $\gamma$-ray flaring episode. The spectral energy distribution of mid-March 2008 is modelled with a homogeneous
   one-zone synchrotron self Compton (SSC) emission plus contributions from
   inverse Compton scattering of external photons from both the accretion disc and
   the broad line region. Indeed, some features in the optical--UV spectrum seem to indicate the
   presence of Seyfert-like components, such as the little blue bump and the
   big blue bump.} 
\authorrunning{F. D'Ammando et al.}
\titlerunning{AGILE detection of a rapid $\gamma$-ray flare from PKS 1510-089 during the
  GASP-WEBT monitoring}
\maketitle
%


%
  \section{Introduction}

Blazars are a subclass of AGN characterized by the emission of strong
non-thermal radiation across the entire electromagnetic spectrum, from radio
to very high energies. Their observational properties include
irregular, rapid and often very large variability, apparent super-luminal
motion, flat radio spectrum, high and variable polarization at radio and
optical frequencies. 
These features are interpreted as the result of
the emission of electromagnetic radiation from a relativistic jet that is
viewed closely aligned to the line of sight (Blandford $\&$ Rees 1978, Urry
$\&$ Padovani 1995). The Spectral Energy Distribution (SED) of blazars
is typically double-humped with a first peak occurring in the IR/optical band
in the so-called $red$ $blazars$ (including Flat Spectrum Radio Quasars, FSRQs, and
Low-energy peaked BL Lacs, LBLs) and at UV/X-rays in the so-called $blue$ $
blazars$ (including High-energy peaked BL Lacs, HBLs). The first peak is
commonly interpreted as synchrotron radiation from high-energy electrons in
a relativistic jet. The second component of the SED, peaking at MeV--GeV energies in
$red$ $blazars$ and at TeV energies in $blue$ $blazars$, is commonly interpreted as
inverse Compton (IC) scattering of seed photons by relativistic electrons
(Ulrich et al., 1997), although a different origin of high energy emission has been
proposed in hadronic models (see e.g., B\"ottcher 2007, for a recent review).

PKS 1510$-$089 is a nearby (z=0.361) radio-loud highly polarized quasar (HPQ)
already detected also in the MeV--GeV energy band by the EGRET instrument on
board $CGRO$ (Hartman et al. 1992). The broadband spectrum of the source
is representative of the class of FSRQ with the inverse Compton component
dominated by the $\gamma$-ray emission and the synchrotron emission peaked
around IR frequencies, even if it is clearly visible in this source a pronunced UV bump
possibly due to the thermal emission from the accretion disk (Malkan $\&$
Moore 1986, Pian $\&$ Treves 1993). 

PKS 1510$-$089 has been extensively observed and studied in the X-ray band by the
satellites EXOSAT (Singh, Rao
$\&$ Vahia 1990, Sambruna et al. 1994), GINGA (Lawson $\&$ Turner 1997),
ROSAT (Siebert et al. 1998), ASCA (Singh, Shrader $\&$ George 1997) and
Chandra (Gambill et al. 2003). The observed X-ray spectrum was very flat in
the 2--10 keV band with photon index of $\Gamma$ $\simeq$ 1.3, but it was steeper ($\Gamma$ $\simeq$ 1.9) in
the ROSAT bandpass (0.1--2.4 keV),
suggesting the presence of a possible spectral break around 1--2 keV. The
difference of the photon indices could be due to the presence of a soft X-ray
excess. Observations by
BeppoSAX (Tavecchio et al. 2000) confirm a possible presence of a soft X-ray excess
below 1 keV. Evidences of a similar soft X-ray excess has been detected in other blazars
such as 3C 273, 3C 279, AO 0235+164 and 3C 454.3, even if the origin of this excess is still an open issue. 

During August 2006, PKS 1510$-$089 was observed in a relatively bright state by $Suzaku$ over approximately
three days and with {\it Swift} 10 times during 18 days as a
Target of Opportunity (ToO) with a total duration of 24.3 ks. $Suzaku$ measured a very hard X-ray spectrum ($\Gamma < 1.5$), which seems to exclude models in which X-rays are produced by synchrotron
radiation of the secondary ultrarelativistic population of electrons and
positrons, as prediced by hadronic models. 
The {\it Swift}/XRT observations instead revealed significant
spectral evolution in the 0.3--10 keV energy band on timescales of one week, with
the spectrum that becomes harder as the source gets brighter (Kataoka et al. 2008). 

Gamma-ray emission from PKS 1510$-$089 was detected several times by EGRET with a integrated flux above 100 MeV between (13 $\pm$ 5) and (49
$\pm$ 18) $\times$ 10$^{-8}$ photons cm$^{-2}$ s$^{-1}$ and an energy spectrum,
integrated over all the EGRET observations, modelled with a power law with photon index
$\Gamma =$ 2.47 $\pm$ 0.21 (Hartman et al. 1999). 

In August 2007, AGILE detected an intense $\gamma$-ray activity from
PKS 1510$-$089. In particular, during the period 28 August
-- 1 September 2007 the average $\gamma$-ray flux observed was F$_{E>100 MeV}$ = ($195 \pm 30) \times 10^{-8}$ photons cm$^{-2}$
s$^{-1}$ (Pucella et al. 2008). Recently, this source was detected again
during high $\gamma$-ray activity states by both the Large Area Telescope (LAT) on board 
the $Fermi$ GST (Ciprini et al. 2008, Tramacere 2009 and Cutini et al. 2009) and AGILE (D'Ammando et
al. 2009, Pucella et al. 2009 and Vercellone et al. 2009). The results of the
AGILE observations during March 2009 will be published in a forthcoming paper (D'Ammando et al., in preparation).

In this paper we present the analysis of the AGILE data obtained during the
observations of PKS 1510$-$089 from 1 March 2008 to 30 March 2008 (see also
D'Ammando et al. 2008a). We also present the radio-to-optical
monitoring of the GASP-WEBT during the period January--April 2008, and the
results of three $Swift$ ToO carried out between 20 and 22 March 2008. 
This broadband coverage over the entire electromagnetic spectrum allows us to
build and study in detail the spectral energy distribution of the source.

The paper is organized as follows. Section 2 describes the AGILE observations, and the
corresponding data analysis. Section 3 introduces the $Swift$
data and the relative analysis.  Section 4 is dedicated to the results of
the GASP-WEBT observations, while in Section 5 we discuss the
spectral energy distribution, its
implication for the emission mechanisms of the source, and finally we draw our conclusions.

Throughout this paper the quoted uncertainties are given at the 1-$\sigma$ level, unless
otherwise stated, and the photon indices are parameterized as $N(E) \propto
E^{-\Gamma}$ (ph cm$^{-2}$ s$^{-1}$ keV$^{-1}$ or MeV$^{-1}$ ) with $\Gamma = \alpha +1$ ($\alpha$ is the
spectral index). We adopt a luminosity distance of d$_L$ = 1915 Mpc for PKS
1510-089, assuming z = 0.361 and a $\Lambda$CDM cosmology with $H_0$ = 71 Km s$^{-1}$
Mpc$^{-1}$, $\Omega_m$ = 0.27 and $\Omega_\Lambda$ = 0.73.
\section{AGILE Data}\label{1510:pointing}
  \subsection{Observation of PKS 1510$-$089} 

The AGILE satellite (Tavani et al. 2008a,b) is an Italian Space Agency (ASI)
Mission devoted to high-energy astrophysics, with four active detectors capable of observing cosmic
sources simultaneously in X-ray and $\gamma$-ray energy bands. 

The Gamma-Ray Imaging Detector (GRID) consists of a combination of a
pair-production Silicon Tracker (ST; Prest et al. 2003, Barbiellini et al. 2001), sensitive in the
energy range 30 MeV--30 GeV, a non-imaging CsI(Tl) Mini-Calorimeter (MCAL; Labanti et al. 2009) sensitive in the 0.3--100 MeV energy band, and a
segmented Anti-Coincidence System (ACS) made of plastic scintillator layers which surrounds all
active detectors (Perotti et al. 2006). A co-aligned coded-mask hard X-ray
imager (SuperAGILE; Costa et al. 2001, Feroci et al. 2007) ensures coverage
in the 18--60~keV energy band. 

The AGILE observations of PKS 1510-089 were performed from 1 March 2008
  12:45 UT to 21 March 2008 2:04 UT, for a total of 211 hours of effective
  exposure time. In the first period, between 1 and 16 March, the source was
  located $\sim 50^\circ$ off the AGILE pointing direction. In the second
  period, between 17 March and 21 March, after a satellite re-pointing, the
  source was located at $\sim 40^{\circ}$ off-axis.
Finally, after a gap of 4 days of observation due to technical maintenance of the
satellite, the source was observed at $\sim$ 50$^\circ$ off axis between 25
March 13:09 UT and 30 March 10:29 UT. Unfortunately during the observation the
  source was substantially off-axis in the field of view of SuperAGILE.


%
\subsection{Data reduction and analysis} \label{1510:dataanal}
AGILE-GRID data were analyzed, starting from the Level--1 data, using the
AGILE Standard Analysis Pipeline (see Vercellone et al. 2008 for a detailed
description of the AGILE data reduction). An ad-hoc
implementation of the Kalman Filter technique is used for track identification
and event-direction reconstruction in detector coordinates and subsequently a
quality flag is assigned to each GRID event, depending
on whether it is recognised as a confirmed gamma-ray event, a charged
particle event, a single-track event, or of uncertain nature. 

After the creation of the event files, the AGILE Scientific Analysis Package
can be run. Counts, exposure, and Galactic background $\gamma$-ray maps were generated with
a bin-size of $0.25^{\circ} \times 0.25^{\circ}$\, for photons with energy
$E>$ 100 MeV. 

To reduce the particle background contamination, we selected only events
flagged as confirmed $\gamma$-ray events, and all events collected during the
South Atlantic Anomaly were rejected. We also rejected all the $\gamma$-ray
events whose reconstructed directions form angles with the satellite-Earth
vector smaller than 80$^{\circ}$, in order to reduce the
$\gamma$-ray Earth albedo contamination.  
The most recent version (BUILD-16) of the Calibration files at the time of writing, publicly available at the ASI Science Data Center
(ASDC) site\footnote{http://agile.asdc.asi.it}, and of the $\gamma$-ray
diffuse emission model (Giuliani et al. 2004) is used.
We ran the AGILE maximum likelihood procedure with a radius of analysis of $10^{\circ}$, on the whole observing
period, in order to obtain the average flux in the $\gamma$-ray band and
  estimate the diffuse parameters used also for measure the daily fluxes, according
  to the procedure described in Mattox et al. (1993). 

\subsection{Results} \label{1510:results}

During the period 2008 March 1--16, AGILE-GRID detected $\gamma$-ray emission
from a position consistent with the powerful $\gamma$-ray quasar PKS 1510$-$089 at a
significance level of 6.2-$\sigma$ with an average flux over the entire period
of $(84 \pm 17) \times
10^{-8}$ photons cm$^{-2}$ s$^{-1}$ for photon energies above 100 MeV.

Instead, in the period 2008 March 17--21, AGILE detected
$\gamma$-ray emission from a position consistent with the source at
a significance level of 7.3-$\sigma$. The AGILE 95$\%$ maximum likelihood
contour level baricentre of the source is $l=351.49^\circ, b=40.07^\circ$,
with a distance between this position and the radio position $(l=351.29^\circ,
b=40.14^\circ)$ of 0.17$^\circ$. 
The overall AGILE error circle, taking both statistical and
systematic effects into account, has a radius $r$ = 0.50$^\circ$.
The average flux above 100 MeV during this
second period, with the source located $\sim 40^\circ$ off the AGILE pointing
direction, was $(134 \pm 29) \times 10^{-8}$ photons cm$^{-2}$ s$^{-1}$. The
peak level of activity with daily integration was $(281 \pm 68) \times 10^{-8}$ photons cm$^{-2}$
s$^{-1}$, showing an increase of a factor two in one day and at least three in
two days, as the source had not been detected for some days after two episodes
of medium intensity. After the sudden increase, the flux rapidly decreased
around March 19, 2008. 

Fitting the data relative to the period March 17--21 with a simple power law model we obtain a photon index of
$\Gamma$ = 1.81 $\pm$ 0.34. This photon index is calculated with the weighted
least squares method, considering for the fit three energy bins: 100--200
MeV, 200--400 MeV and 400--1000 MeV. The photon index obtained for this
second period is consistent within the errors with the one observed by AGILE in August 2007
($\Gamma$ = 1.98 $\pm$ 0.27). 

Figure 1 shows the $\gamma$-ray light-curve between 1 and 21 March 2008 with 2-day resolution for the first
period and 1-day for the second period, for photons of energy above 100
MeV. The downward arrows represent 2-$\sigma$ upper limits.  Upper limits are calculated when the analysis
provides a significance of detection $<$ 3-$\sigma$ (see Mattox et al. 1996).

Finally, in the third period between 25 and 30 March 2008 the source was not detected by
the GRID and an upper limit with 95$\%$ confidence level of 54  $\times
10^{-8}$ photons cm$^{-2}$ s$^{-1}$ is provided.   

During August--October 2008, $Fermi$-LAT detected the
source with an average flux for E $>$
100 MeV of (55.8 $\pm$ 3.3) $\times 10^{-8}$ photons cm$^{-2}$ s$^{-1}$ and a
peak of intensity of (165.9 $\pm$ 11.7) $\times 10^{-8}$ photons cm$^{-2}$ s$^{-1}$ (Abdo et
al. 2009). The peak of $\gamma$-ray emission corresponds to the first flare
observed by $Fermi$-LAT at the end of September 2008 (Tramacere 2008); the average flux value confirms the flaring state observed by
AGILE in mid-March.  

Moreover in August--October 2008, $Fermi$-LAT observed a softer
photon index for this source, $\Gamma$ = 2.48$\pm$0.05
(Abdo et al. 2009), but
this value corresponds to an average value over three months of
observation in which the source flux was variable, whereas the
value reported by us refers to a rapid flaring episode. The value
obtained by $Fermi$-LAT is very similar to
that measured by EGRET averaging over all the observations ($\Gamma$ = 2.47
$\pm$ 0.21), confirming that the average spectral indexes are softer than
those measured during short flaring states. The
  difference between the value obtained by AGILE and $Fermi$ could also be partially due to the different bandpasses of the
  two intruments.
  
\begin{figure}
 \centering
 \includegraphics[width=9.5cm]{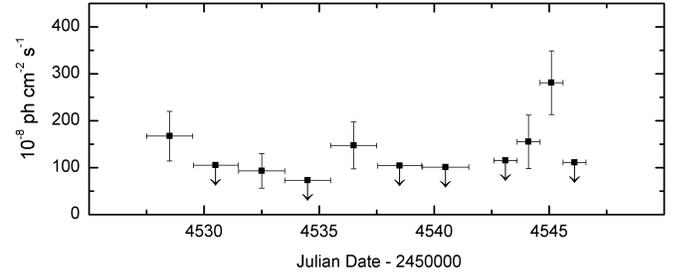}
   \caption{AGILE-GRID $\gamma$-ray light curve between 1 and 21 March 2008 at 1-day or 2-day resolution for E
            $>$ 100 MeV with fluxes in units of 10$^{-8}$ photons cm$^{-2}$
            s$^{-1}$. The downward arrows represent 2-$\sigma$ upper limits.}
   \label{1510GRIDlc}
\end{figure}
\section{SWIFT observations}

The NASA $Swift$ gamma-ray Burst Mission (Gehrels et al. 2004), performed
three ToO observations of PKS 1510--089 in three consecutive days with the
first occurring on 20 March 2008. The three
observations were performed using all three on-board experiments: the X-ray
Telescope (XRT; Burrows et al. 2005, 0.2--10 keV), the UV and Optical Telescope
(UVOT; Roming et al. 2005, 170--600 nm) and the coded-mask Burst Alert
Telescope (BAT; Barthelmy et al. 2005, 15--150 keV). The hard X-ray flux of
this source is below the sensitivity of the BAT instrument for so short exposure
and therefore the data from this instruments will not be used.

\begin{table*}[th!]
\caption{Observation log and fitting results of $Swift$/XRT observations of PKS 1510-089. Power law model with $N_{\rm
H}$ fixed to Galactic absorption. $^{a}$ Unabsorbed flux. $^{b}$ Cash
  statistic (C-stat) and percentage of Montecarlo realizations that had
  statistic $<$ C-stat, performing 10$^4$ simulations.}
\centering
\begin{tabular}{cccccc}
\hline
\hline
\noalign{\smallskip}
\multicolumn{1}{c}{Observation} &
\multicolumn{1}{c}{Exposure Time} &
\multicolumn{1}{c}{Counts} &
\multicolumn{1}{c}{Flux 0.3-10 keV$^{a}$} &
\multicolumn{1}{c}{Photon index} &
\multicolumn{1}{c}{$ \chi^{2}_{red}$ (d.o.f.) / $C$-$stat$ ($\%$)$^{b}$} \\
\multicolumn{1}{c}{Date} &
\multicolumn{1}{c}{ (sec) } &
\multicolumn{1}{c}{ (0.2 - 10 keV) } &
\multicolumn{1}{c}{erg cm$^{-2}$ s$^{-1}$} &
\multicolumn{1}{c}{$\Gamma$}&
\multicolumn{1}{c}{} \\
\hline
\noalign{\smallskip}
20-Mar-2008 & 1961 & 306 & $12.20 _{-1.65}^{+1.65} \times 10^{-12}$ & $1.16 \pm 0.16$ & 0.87 (13)\\
21-Mar-2008 & 1966 & 189 & $8.77 _{-1.39}^{+1.33} \times 10^{-12}$ & $1.53 \pm 0.17$ &  467
(48.2)$^{b}$\\
22-Mar-2008 & 1885 & 261 & $9.48 _{-1.14}^{+1.13} \times 10^{-12}$ & $1.41 \pm 0.19$ & 1.34
(11) \\
\noalign{\smallskip}
\hline
\hline
\end{tabular}
\label{SWIFT_tab}
\end{table*}

\subsection{Swift/XRT Data}

The XRT data were processed with standard procedures ({\tt xrtpipeline}
v0.12.0), the filtering and screening criteria were applied by means of the FTOOLS in the {\tt Heasoft} package v6.5. 
Given the low rate of PKS 1510--089 during the three observations ($<$ 0.5
count s$^{-1}$ in the 0.2--10 keV range), we only considered photon counting
(PC) data for our analysis, and further selected XRT grades 0--12 (according to $Swift$
nomenclature, see Burrows et al. 2005). 
No pile up correction was necessary.
The ancillary response files were generated with the task {\tt xrtmkarf}, applying
corrections for the PSF losses and CCD defects, and we
used the latest spectral redistribution matrices (RMF, v011) in the
Calibration Database maintained by HEASARC. 
The adopted energy range for spectral fitting is 0.3--10 keV and all data were
rebinned with a minimum of 20 counts per energy bin to use the
$\chi^{2}$ minimization fitting technique.  
An exception is the observation of 2008 March 21,
when the number of counts was so low that the Cash statistic (Cash, 1979)
on ungrouped data was used. $Swift$/XRT uncertainties are given at 90$\%$ confidence
level for one interesting parameter, unless otherwise stated.
 
Spectral analysis was performed using the {\tt XSPEC} fitting package 12.4.0
(Arnaud et al. 1996). We fitted the spectra with a power law model with Galactic absorption fixed to
$N_{\rm_H}$ = 6.89 $\times$ 10$^{20}$ cm$^{-2}$ (Kalberla et al. 2005). 
Table 1 summarizes the most important information on XRT observations and the
relative spectral fit parameters.

A variability of about 30$\%$ in the X-ray flux of the source was observed on time scale of
one day.
Notwithstanding the uncertainties due to the errors on fluxes and photon
  indexes, the XRT data seem to indicate that
the X-ray spectrum becomes harder when the source gets brighter, confirming the behaviour
already observed in this source by Kataoka et al. (2008) during the $Swift$/XRT
observations carried out in August 2006. This is a trend often
observed in HBL (see e.g Massaro et al. 2008, Tramacere et al. 2007, Kataoka
et al. 1999), but quite rare for quasar-hosted blazars
such as PKS 1510$-$089. For instance, 3C 454.3 shows approximately the same
spectral slope in different brightness states (see e.g. Raiteri et al. 2007;
Raiteri et al. 2008).

\subsection{Swift/UVOT Data}

During the three $Swift$ pointings, the UVOT instrument (Poole et al.
2008) observed PKS 1510$-$089 in all its optical ($V$, $B$, and $U$) and UV (UV$W1$,
UV$M2$, and UV$W2$) photometric bands. Data were reduced with the {\tt uvotmaghist} task
of the HEASOFT package. Source counts were extracted from a circular region of 5 arcsec
radius, centred on the source, while the background was estimated from a surrounding
annulus with 8 and 18 arcsec radii. In the first two days only one exposure per filter
was available, while three exposures per filter were acquired in the last day. With the
only exception of UV$M2$, the source brightness resulted quite stable in all the UVOT
bands, with variations of a few hundredth of mag, well inside the typical UVOT data
uncertainty of 0.1 mag due to both systematic and statistical errors. Average values are:
$V=16.94$, $B=17.19$, $U=16.31$, UV$W1=16.64$, and UV$W2=16.55$. The UV$M2$ frames have
low signal-to-noise ratios, thus the source magnitude in this band presents a larger
dispersion; the average value is UV$M2=16.47 \pm 0.14$.


\section{Radio-to-optical observations by the GASP}

PKS 1510--089 is one of the 28 $\gamma$-ray-loud blazars that are regularly monitored by
the GLAST-AGILE Support Program (GASP; Villata et al. 2008) of the Whole Earth Blazar
Telescope (WEBT). Optical and near-IR data are collected as already calibrated
magnitudes, according to a common choice of photometric standards from Raiteri
et al. (1998). Radio data are
provided as calibrated flux densities.
The reference optical band for the GASP is the $R$ band; the corresponding light curve in
January--April 2008 is shown in the top panel of Fig. 1, with the data
provided by the following observatories:
Abastumani, Calar Alto\footnote{Calar Alto data were acquired as part of the MAPCAT
(Monitoring AGN with Polarimetry at the Calar Alto Telescopes) project.}, Crimean, Lowell
(Perkins), Lulin, Roque de los Muchachos (KVA and Liverpool), San Pedro Martir, St. Petersburg,
Torino.
The source showed intense activity during all the considered period, with several
episodes of fast variability.
At the beginning of the optical observing season, the January observations indicate that the source was in a faint state, around $R=16.6$. A fast brightening of
$\sim 1.3$ mag in 8 days led the source to $R=15.3$ on February 15. This was followed by
a $\sim 0.6$ mag dimming in 4 days. Other minimum brightness states were observed on
March 23 and in late April, while peaks were detected on March 29 and April 11.
Near-IR data in the $JHK$ bands were taken at Campo Imperatore and Roque de los Muchachos (Liverpool).
Millimetric flux densities at 345 and 230 GHz came from the Submillimeter Array (SMA) on
Mauna Kea. Centimetric radio data were acquired at Medicina (22 and 8 GHz), Noto (43
GHz), and UMRAO (14.5, 8.0, and 4.8 GHz). In Fig. 1 the source radio behaviour
in different bands is compared to the optical one (top panel). The light
curve at high radio frequencies (230--345 GHz) suggests that the mechanism producing the flaring events observed in the optical band in 
the second half of February and in late March--April 2008 also interested the millimetric 
emitting zone, with some delay. An estimate of this delay is hampered by the
limited data sampling. At lower radio frequencies (22--43 GHz) a hint of flux increase is 
visible in the second part of the light curve,
while the radio flux at 5--15 GHz shows no trend.
This suggests that the jet regions that are responsible for the emission at the longest 
radio wavelengths are not affected by the flaring mechanism.

\begin{figure*}[th!]
\sidecaption
\includegraphics[width=11.5cm]{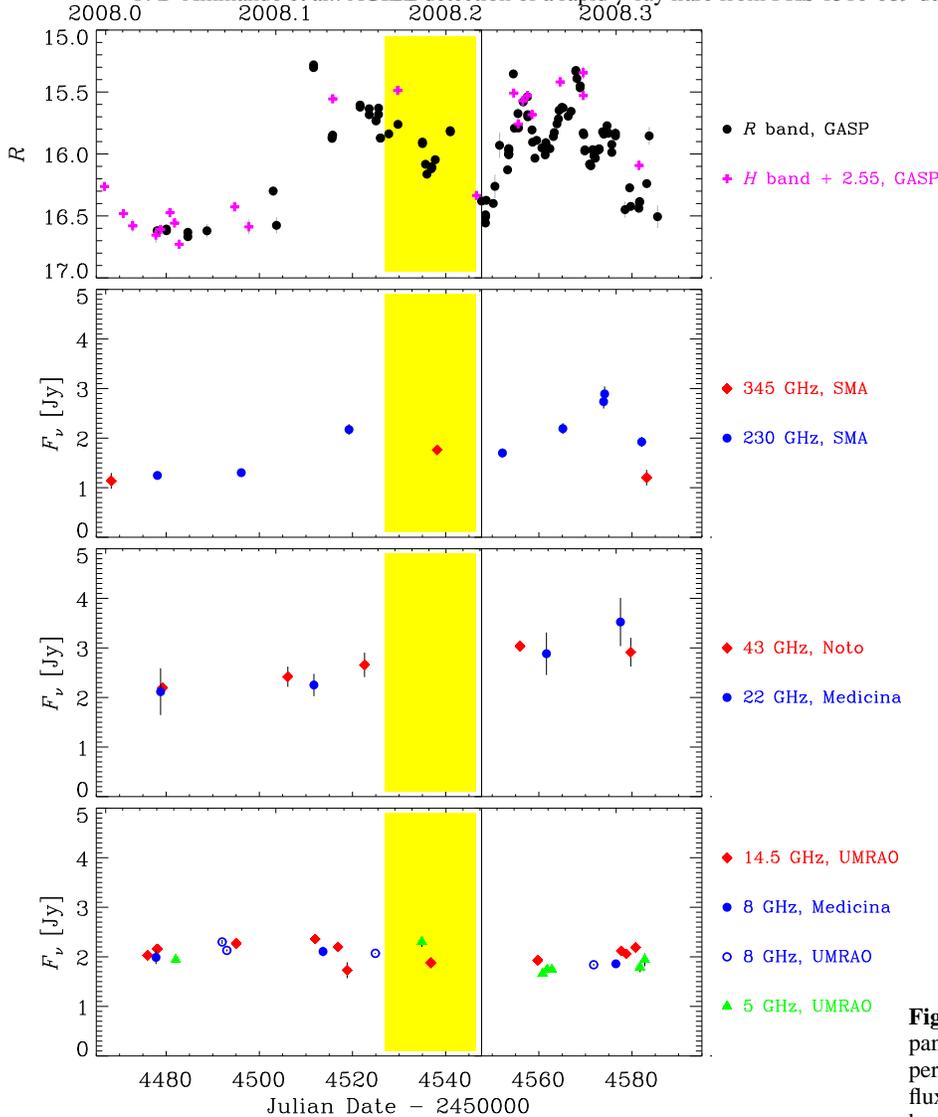}
  \caption{Optical light curve of PKS 1510-089 (top panel) obtained by the
    GASP--WEBT during the period January-April 2008 compared to its radio flux densities at different frequencies.
The vertical bar indicates the time of the $Swift$ observations. The yellow
shaded region marks the period also covered by the AGILE observation.}
\end{figure*}

\section{Discussion and conclusions} \label{1510:discussion}

In the last two years, PKS 1510$-$089 showed a very high activity in $\gamma$-ray band with several flaring episodes (see Pucella et al. 2008,
Tramacere 2008, Ciprini et al. 2009, D'Ammando et al. 2009, Pucella et
al. 2009, Vercellone et al. 2009 and Cutini et al. 2009).
During the period January--April 2008, the source showed a high variability over all
the electromagnetic spectrum from radio to $\gamma$-rays, with several flaring
episodes in the optical band and a rapid and intense flare detected in the $\gamma$-ray
band in mid-March. 

\begin{figure*}
\vspace{-0.4cm}
 \centering 
  \includegraphics[width=14cm, height=10cm] {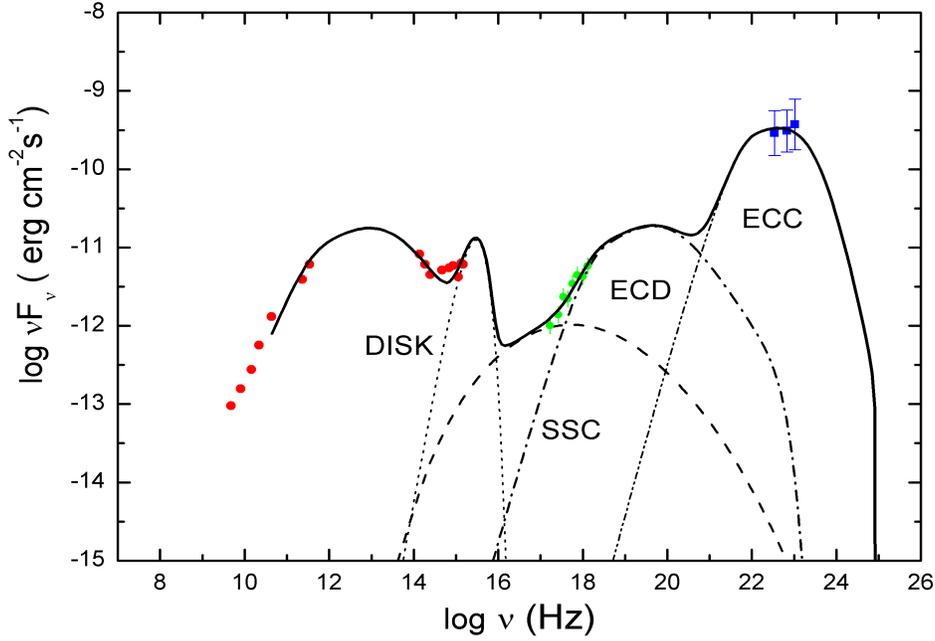}
\vspace{-0.4cm}
   \caption{Spectral Energy Distribution of PKS 1510-089 for the AGILE-GRID
            observation of 17--21 March 2008, 
            including quasi-simultaneous GASP radio-to-optical data, the $Swift$/UVOT data of 20--22 March and the $Swift$/XRT data of 20
            March. The dotted, dashed, dot--dashed and double--dot--dashed
            lines represent the accretion disk black body, the SSC,
            the ECD and the ECC radiation, respectively.} 
  \label{1510_SED}
\end{figure*}

\subsection{Modelling the Spectral Energy Distribution}

Figure \ref{1510_SED} shows the spectral energy distribution for the AGILE-GRID
observation period 2008 March 17--21, including quasi-simultaneous optical and radio data by GASP, and UVOT and X-ray data by $Swift$.
Since the source brightness over the three days of UVOT observations remained stable, we
built a unique SED for the whole period, including
contemporaneous data at other frequencies, in particular radio-to-optical data from the
GASP. The optical and near-IR data were acquired exactly in the period of the UVOT
observations: one $R$-band datum from Roque (KVA) and $J$, $H$, and $K$ data from Campo
Imperatore.
The UVOT and GASP magnitudes were corrected for Galactic extinction by adopting
$A_B=0.416$ mag, and deriving the values in the other bands according to Cardelli et al.
(1989). To convert magnitudes into fluxes, we assumed the zero-mag fluxes by Poole et al.
(2008) and Bessel et al. (1998). 

Radio light curves are less well-sampled than the
$R$-band one (see Fig. 2); the high-frequency radio data (43--345 GHz) shown in the SED were taken
within a week from the UVOT observations, while the low-frequency data points were
obtained by interpolating between the closest data preceeding and following the UVOT
observations. This is justified by the smooth behaviour of the low-frequency radio light
curves.

The dip in the SED corresponding to the UV$W1$ frequency must be regarded with cautious,
since it is also found for other blazars with different redshift and could be systematic. 
Observations performed by $Swift$ and the GASP in March and June 2007, when the source was
nearly at the same brightness level, showed the same shape in the near-IR-to-UV
part of the SED. A similar trend characterized the optical--UV SED of August
2006 shown by Kataoka et al. (2008). We notice that the shape of the SED in the optical band may be affected by the flux
contribution of broad emission lines, including the little blue bump (Neugebauer et al.
1979, Smith et al. 1988).

For the SED we used the
$Swift$/XRT data collected on 20 March, the observation closest to the $\gamma$-ray
flare and during which the higher X-ray flux was observed.

In order to model the spectral energy distribution we used a homogeneous
one-zone synchrotron self Compton (SSC; Marscher $\&$ Gear 1985, Maraschi
et al. 1992, Bloom $\&$ Marscher 1996)
model, plus the contribution of external Compton scattering of both direct disk
radiation (ECD; Dermer et al. 1992) and photons from
the broad line region (BLR) clouds (ECC; Sikora et al. 1994). 
The strong thermal features usually observed in FSRQs (and in this blazars in
particular, see Neugebauer et al. 1978, Smith et al. 1988) at optical/UV frequencies suggest
that the environment is rich of soft photons produced by the accretion disk
and/or reprocessed by the BLR. This implies that the energy density of the
external soft radiation is much higher than that of the synchrotron radiation,
therefore during the $\gamma$-ray flares in FSRQs the most important processes
are the ECC and ECD and the $\gamma$-ray photon index could be determined
  by the dominant contribution of the two.
  
We consider a moving spherical blob of radius $R$, filled by relativistic
electrons and embedded in a random magnetic field. We assume that the electron energy density distribution is described by
a broken power law:
\begin{equation}
n_{e}(\gamma)=\frac{K\gamma_{b}^{-1}}{(\gamma /\gamma_{b})^{p_1}+(\gamma
/\gamma_{b})^{p_2}}
\end{equation}
where $\gamma$ is the electron Lorentz factor assumed to vary between
$\gamma_{min}$ and $\gamma_{max}$, 
{\it p$_1$} and {\it p$_2$} are pre- and post-break
electron distribution spectral indexes, respectively, and $\gamma_b$ is the
break energy Lorentz factor. We assume that the blob
contains a comoving random average magnetic field B with a bulk Lorentz factor
$\Gamma$ at an angle $\theta$ with respect to the line of sight. The
relativistic Doppler beaming factor is then $\delta$ =
[$\Gamma$(1-$\beta$cos$\theta$)]$^{-1}$, and $K$ is the normalization density parameter
into the blob. 

We have chosen an angle of view of 0.05 rad in agreement with both the apparent
jet velocities derived from multiepoch Very Long Baseline Array (VLBA)
observations of the source
(Homan et al. 2001; Wardle et al. 2005; Jorstad et al. 2005; Lister et
al. 2009) and the value used by Kataoka et al. (2008). 

\begin{table}
\begin{center}
\begin{tabular}{c c r} 	
\hline  
\hline
\noalign{\smallskip}
 Parameter & Value & Units \\ 
  \hline
\noalign{\smallskip}
  $p_{\rm 1}$             & 2.2   & \\
  $p_{\rm 2}$             & 4.6   & \\
  $\gamma_{\rm min}$           & 30    & \\
  $\gamma_{\rm b}$             & 290   & \\
  $\gamma_{\rm max}$           & 5200  & \\
  $K$                          & 75    & cm$^{-3}$ \\
  $R$                          & 10    & 10$^{15}$\,cm\\
  $B$                          & 3.5    & G\\
  $\delta$                     & 20.26 & \\
  $L_{\rm d}$                  & 5     & 10$^{45}$\,erg\,s$^{-1}$\\
  $\theta$                 & 2.86   & degrees\\
  $\Gamma$                     & 18   & \\
\noalign{\smallskip}
\hline
\hline
\end{tabular}
\caption{Parameters for the model used to explain the SED of PKS 1510-089
  during the $\gamma$-ray flare of March 18-19, 2008.}
\end{center}
\end{table} 

The short time variability observed in $\gamma$-ray band constrains the size of the emitting region to $R$ $<$ c$\Delta$t$_{var}$$\delta$ /(1+z) = 3.86
$\times$ 10$^{16}$ cm, where $\Delta$$t_{var}$ is the observed variation time.
An accretion disk characterized by a black
body spectrum with a luminosity of 5 $\times$ $10^{45}$ erg s$^{-1}$, as
estimated with UV observations by Pian $\&$ Treves (1993), at 0.05 pc from the
blob is assumed as one of the sources of external target photons. We also assumed
a BLR at 0.2 pc, reprocessing 10$\%$ of the irradiating continuum.
The IC spectra derived from the approximation of the BLR radiation as a black body reproduces quite well more refined spectra
calculated by Tavecchio et al. (2008) taking into account a more accurate
shape of the BLR.  

Assuming a model with synchrotron, SSC and EC components plus the contribution from the
 accretion disk radiation, the spectral energy distribution of mid March 2008
  can be well represented with input parameters
  summarized in Table 2. In the choice of the parameters we were guided
  by the knowledge of the angle of view, the disc luminosity and the
  simultaneous observations of the synchrotron and IC peak regions (therefore of the
  synchrotron and IC peak frequencies and luminosities). In addition, the minimum variability timescale
  gives an indication of the size of the emitting region. However, even if these quantities are quite well tightly
  contrained by fitting the whole SED, the choice of some parameters is not
  unique because the contemporaneous presence of the synchrotron, SSC, ECC and ECD
  components leads to a possible partial degeneration of the parameters.

\subsection{X-ray spectral evolution}

The spectral evolution detected in X-rays by $Swift$ in just two days, soon after the
$\gamma$-ray flaring episode is another hint of the rapid change of activity of this source. Usually in FSRQs as PKS 1510--089 only little
variability is observed in X-rays on short timescales from hours to days, and
also on longer timescales the X-ray spectral shape
is almost constant with only little variations. The photon indices measured
  by $Swift$/XRT in March 2008, in particular during the first observation,
  tend to be lower than usually observed in FSRQs ($<$$\Gamma$$_{[0.1-2.4]}$$>$ = 1.76 $\pm$ 0.06
  and $<$$\Gamma$$_{[2-10]}$$>$ =
1.65 $\pm$ 0.04, Donato et al. 2001) and are more similar to those observed
in some high redshift
quasars (such as RBS 315, Tavecchio et al. 2007, and Swift J0746+2548, Sambruna et
al. 2007). The hard photon indices of high redshift blazars
could be interepreted in terms of absorption by warm plasma in the region surrounding
the source, in agreement with a scenario where in the early evolution phases the quasars are substantially obscured by gas subsequently
expelled from the host galaxy by powerful winds (see e.g. Fabian 1999). However, considering the low redshift, this interpretation
is unlikely for this source.

Instead the X-ray spectral evolution observed by $Swift$/XRT could be due to the contamination of an
additional component below $\sim$2 keV: the soft
X-ray excess. In fact previous observations with Chandra (Gambill et
  al. 2004) and Suzaku (Kataoka et al. 2008) seem to indicate the presence of
  the soft X-ray excess in the spectrum of PKS 1510--089. The soft X-ray excess is an emission in excess of the
extrapolation of the power law component dominating at higher energies, but
the origin of this excess in AGNs is still an open issue  (see e.g. D'Ammando
et al. 2008b for a detailed discussion). In the past, it was often associated with the
thermal emission of the accretion disc and then related to the big blue
bump. However, it has recently been shown that modelling the soft X-ray excess
in non-blazar AGNs with a
thermal component yields a disc temperature remarkably constant,
around 0.1--0.2 keV, regardless the central black hole mass and luminosity
(Gierlinsky $\&$ Done 2004, Crummy et al. 2006). Also in Kataoka et al. (2008)
the soft X-ray excess is tentatively described by a black body with
temperature kT $\simeq$ 0.2 keV. This result is difficult to
explain in any model for the soft excess related to disc continuum emission,
as in any disc model the temperature is expected to vary with both the black
hole mass and the accretion rates. 

For FSRQs one possible theoretical
explanation is that the soft X-ray excess is a bulk Comptonization feature
produced by cold plasma accelerated in a jet (Celotti et al. 2007), even if
until now this feature has never been positively observed; in BL Lac objects instead
  the radiative environment is too weak to produce the soft X-ray excess via
  bulk Compton and the soft X-ray excess is likely related to the high energy
tail of the synchrotron emission. 

The change of photon index observed during the $Swift$/XRT observations could be due to the fact that the
spectral shape of the inverse Compton component in X-ray remains roughly
constant, but the amount of contamination from the soft excess emission
varies. The contribution of the soft X-ray excess would be more important when the source gets fainter, affecting the spectrum at higher energies. Unfortunately the brief
exposure of the $Swift$ observation does not allow a detailed spectral
modelling of this feature.

A possibility for the origin of this hard power law in PKS 1510-089 is that the photon index observed in
X-rays is due to the combination of the synchrotron self Compton and external
Compton emission and therefore due to the mismatch of the spectral slopes of these
two components, not to the presence of a real soft X-ray excess. This is
the solution that the data presented in this paper would favour. In this
context, the spectral evolution during the three $Swift$/XRT observations could
be due to the change of contribution of one of the two components and
therefore to a different variability of the SSC and EC components.

\begin{figure}
 \centering 
 \includegraphics[width=8.0cm]{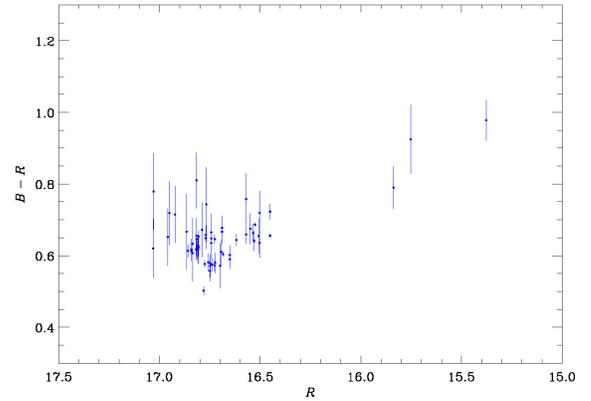}
   \caption{$B-R$ color index versus $R$-band magnitude for PKS 1510--089 obtained with archive
   data of the Torino Observatory.}
  \label{1510_B_R}
\end{figure}

\subsection{Thermal emission components}

Even if the SED of the blazars are usually dominated by the
beamed non-thermal jet radiation, some of them show the signature of
features Seyfert-like as the little blue bump and the big blue bump. The
little blue bump is usually observed in quasars between $\sim$ 2000 and $\sim$
4000 $\AA$ in the rest frame and it is likely due to the contribution of FeII
and MgII emission lines and the Balmer continuum produced in the Broad Line
Region (Wills et al. 1985). The big blue bump instead is associated with a
rise in the UV band commonly interpreted as thermal emission from the
accretion disc (see e.g. Laor 1990). Evidences of these thermal components have
been found in other quasar-like blazar as 3C 273 (Grandi and Palumbo 2004;
T\"urler et al. 2006), 3C 279 (Pian et al. 1999) and 3C 454.3 (Raiteri et al. 2007).  

The presence of the emission by the accretion disk is
consistent with the scenario in which the seed photons for the IC producing
the $\gamma$-rays are external to the relativistic jet but usually it is not
observed because hidden by the beamed variable synchrotron emission. The fact
that the synchrotron component of PKS 1510$-$089 peaks around
10$^{13}$ Hz (see Bach et al. 2007, Nieppola et al. 2008) allows us to
observe these thermal features in this source. In fact around 10$^{15}$ Hz a rising emission is
visible in the spectrum and it is likely a manifestation of the big blue bump produced by the
accretion disk, as already discussed for this source by Malkan $\&$ Moore (1986) and Pian $\&$
Treves (1993); moreover a hint of the presence of the little blue bump seems
to appear in the SED of the source at 10$^{14.5}$ Hz. 

Given the redshift of PKS 1510--089, 
the H$\alpha$, H$\beta$, FeII and MgII lines mostly contribute to the observed
spectrum between 10$^{14.2}$ and 10$^{14.8}$ Hz, and together with the
disc emission, could explain the excess of emission observed around
10$^{14.5}$ Hz and not modelled from the other components represented in the
SED. Moreover the presence of these non-jet components in the blue part of
the spectrum of this blazar has already been observed by Neugebauer et al. (1979)
and Smith et al. (1988) and it is in agreement with the redder-when-brighter behaviour shown by
the $B-R$ index versus $R$-band plot of Fig.\ 4. The plot has been obtained with archive
data stored at the Torino Observatory. 

\subsection{Energetics and alternative model}

Finally to estimate the energetics of PKS 1510--089 we compute the isotropic
luminosity in the $\gamma$-ray band, comparing with the Eddington and
bolometric luminosity and the total power transported by the jet. 
For a given source with redshift $z$, the isotropic emitted
luminosity in the energy band $\epsilon$ is defined as:
\begin{equation}
  \label{eq_lum}
  L(z)_{\epsilon} = \frac{4\pi F d_{\rm l}^{2}(z)}{(1+z)^{(1-\alpha_{\gamma})}} \,,
\end{equation}
where, in our case, $\epsilon$ is the $\gamma$-ray energy band
with ${E_{\rm min}}=100$\,MeV and  ${E_{\rm max}}=10$\,GeV,
$\alpha_{\gamma}$ = $\Gamma$ -- 1, $F$ is the $\gamma$-ray energy flux between
${E_{\rm min}}$ and  ${E_{\rm max}}$ calculated from the photon flux
$F_{\gamma}$ ($E >$ 100 MeV) as suggested by Ghisellini et al (2009):

\begin{equation}
  F= 1.6 \times 10^{-12} \frac {\alpha_{\gamma} F_{\gamma}}{1-\alpha_{\gamma}} [100^{1-\alpha_{\gamma}}-1]
\end{equation}
The luminosity distance is given by
\begin{equation}
  \label{dist_lum1}
  d_{\rm l}(z_1,z_2)= (1+z_2)^{2} \times \frac{c/H_0}{1+z_2}\int_{z_1}^{z_2}
  [E(z)]^{-1} dz\,,
\end{equation}
where $\rm z_1 = 0$, $\rm z_2 = z_{src}$ and
\begin{equation}
  E(z)=\sqrt{\Omega_{\rm m} (1+z)^3
    +(1-\Omega_{\rm m}-\Omega_{\Lambda})(1+z)^2+\Omega_{\Lambda}}\,,
\end{equation}
where $H_0$ is the Hubble constant,  $\Omega_m$  and $\Omega_{\Lambda}$ are
the contribution of the matter and of the cosmological constant, respectively, to
the density parameter. Using a luminosity distance  d$_L$ = 1915 Mpc and the
average $\gamma$-ray flux observed by the AGILE-GRID during 17--21 March 2008, we obtain for PKS
1510--089 (z = 0.361) an isotropic luminosity $L_{\gamma}^{\rm iso}$ = 5.3 $\times$ $10^{47}$\, erg s$^{-1}$.

The power carried by the jet in the form of magnetic field ($L_B$), cold protons ($L_p$), relativistic electrons ($L_e$) and produced
radiation ($L_{rad}$), are:

\begin{equation}
  \label{lum_p}
  L_{\rm p} = \pi\,R^{2}\,\Gamma^{2}\,c\,\int[N(\gamma)\,m_p\,c^2\,d\gamma] =
  3.6 \times 10^{45}\, \textrm {erg s$^{-1}$}
\end{equation}

\begin{equation}
  \label{lum_e}
  L_{\rm e} =
  \pi\,R^{2}\,\Gamma^{2}\,c\,\int[N(\gamma)\,\gamma\,m_e\,c^2\,d\gamma] = 1.5
  \times 10^{44}\, \textrm{erg s$^{-1}$}
\end{equation}

\begin{equation}
  \label{lum_B}
  L_{\rm B}  = \pi\,R^{2}\,\Gamma^{2}\,c\,U_{B} = 1.5 \times 10^{45}\, \textrm{erg s$^{-1}$}
\end{equation}

\begin{equation}
  \label{lum_r}
  L_{\rm rad}  \simeq L_{\rm iso}\Gamma^{2}\,/\delta^4  = 5.3 \times 10^{45}\,
  \textrm{erg s$^{-1}$}
\end{equation}

where $U_B$ is the magnetic energy density. Therefore the total power transported by the jet is $P = L_B + L_p + L_e +
L_{rad}$ = 1.1 $\times$ 10$^{46}$ erg s$^{-1}$. 

Assuming for the source a black hole mass $M_{\rm BH}$ = 4.5 $\times$ 10$^{8}$ M$_{\odot}$
(Woo $\&$ Urry 2002), we obtain an Eddington luminosity 

\begin{equation}
L_{\rm Edd} = \frac{4\pi\,G\,c\,m_H\,}{\sigma_T} M_{BH} = 5.7 \times 10^{46}\,
\textrm{erg s$^{-1}$}
\end{equation}

to be compared with the bolometric luminosity $L_{bol}$ = 2.4 $\times$
10$^{46}$ erg s$^{-1}$ reported in Woo $\&$ Urry (2002).
 
An alternative theoretical model was recently proposed by Kataoka et
al. (2008) to interprete the data of PKS 1510--089 collected during August
2006, with the high energy emission originated by the Comptonization of infrared
radiation produced by the molecular torus surrounding the central engine and
suggesting that the soft X-ray excess
could be produced by the IC scattering of external photons by a population
of cold electrons, as discussed by Begelman et al. (1997) and Celotti et al. (2007).
An accurate theoretical interpretation of the SED is beyond the scope of this
paper; however, our data seem not to rule out this alternative model. Infact,
we have not simultaneous observations in FIR band that can confirm the excess
detected by IRAS (Tanner et al. 1996), interpreted by Kataoka et al. as due to
dust radiation from the nuclear torus and assumed as main source of seed photons
for the IC mechanism. Moreover, the bulk Comptonization feature should not
be easily observable during a high activity state of the source, such as that
observed in mid-March 2008, because overwhelmed by the SSC and ECD emission
and therefore this is not the best situation to test this hypothesis. 

Further X-ray observations with XMM-$Newton$ and $Suzaku$, simultaneously with
the optical monitoring by means of REM Telescope and WEBT Consortium, could give important
indication of the emission mechanisms involved in this source, in particular of the
real nature of the soft X-ray excess, the presence in the spectrum of Seyfert-like features
and the existence of the bulk Comptonization feature or not. 

Finally, with two $\gamma$-ray satellites, AGILE and $Fermi$, in orbit at the same time we will be
able to study in detail the source behaviour at high energies on long time
scale, even if a wide multiwavelength coverage is necessary to
achieve a complete understanding of the structure of the jet, the origin of
the seed photons for the inverse Compton process and all the emission
mechanisms working in this blazar.

\begin{acknowledgements}

We thank the anonymous referee for the useful comments.
The AGILE Mission is funded by the Italian Space Agency (ASI) with scientific
and programmatic participation by the Italian Institute of Astrophysics (INAF)
and the Italian Institute of Nuclear Physics (INFN). 
We acknowledge the use of
public data from the Swift data archive. We thank Swift Team for making these
observations possible, particularly the duty scientists and science planners.
The Submillimeter Array is a joint project between the Smithsonian
Astrophysical Observatory and the Academia Sinica Institute of Astronomy and
Astrophysics and is funded by the Smithsonian Institution and the Academia
Sinica. UMRAO is funded by a series of grants from the NSF and by the University of Michigan. The research has been
supported by the Taiwan National Science Council grant No. 96-2811-M-008-058. This paper is
partly based on observations carried out at the German-Spanish Calar Alto Observatory,
which is jointly operated by the MPIA and the IAA-CSIC. Acquisition of the MAPCAT
data is supported in part by the Spanish ``Ministerio de Ciencia e Innovaci\'on" through
grant AYA2007-67626-C03-03. Some of the authors acknowledge financial support by the Italian Space Agency through
contract ASI-INAF I/088/06/0 for the Study of High-Energy Astrophysics.

\noindent {\it Facilities}: AGILE, $Swift$, UMRAO and WEBT.

\end{acknowledgements}
%
%

\end{document}